\documentclass[aps,prb,showpacs,showkeys,preprint]{revtex4-1}
\usepackage{amssymb}
\usepackage{amsmath}
\usepackage{amsfonts}
\usepackage{graphicx}%
\usepackage{epstopdf}
\usepackage{sidecap}
\sidecaptionvpos{figure}{c}
\usepackage[note-name=, use-sort-key = false]{notes2bib}
\usepackage{color}

\begin{document}
\preprint{HEP/123-qed}
\title{Optical spectroscopy study on pressure-induced phase transitions in the three-dimensional Dirac semimetal Cd$_3$As$_2$}

\author{E. Uykur$^{*,1}$, R. Sankar$^{2}$, D. Schmitz$^{3}$, and C. A. Kuntscher$^{\dagger,1}$}

\affiliation{$^1$Experimentalphysik 2, Universit\"{a}t Augsburg, D-86159 Augsburg, Germany}

\affiliation{$^2$Institute of Physics, Academia Sinica, Taipei, 10617, Taiwan}

\affiliation{$^3$Chemical Physics and Materials Science, Universit\"{a}t Augsburg, D-86159 Augsburg, Germany}


\begin{abstract}

We report a room-temperature optical reflectivity study performed on [112]-oriented Cd$_3$As$_2$ single crystals over a broad energy range under external pressure up to 10 GPa. The abrupt drop of the band dispersion parameter ($z$-parameter) and the interruption of the gradual redshift of the bandgap at $\sim$4~GPa confirms the structural phase transition from a tetragonal to a monoclinic phase in this material. The pressure-induced increase of the overall optical conductivity at low energies and the continuous redshift of the high-energy bands indicate that the system evolves towards a topologically trivial metallic state, although a complete closing of the band gap could not be observed in the studied pressure range. Furthermore, a detailed investigation of the low-pressure regime suggests the possible existence of an intermediate state between 2 and 4~GPa , that might be a precursor of the structural phase transition or due to the lifted degeneracy of the Dirac nodes. Several optical parameters show yet another anomaly at 8~GPa, where low-temperature superconductivity was found in an earlier study.

\end{abstract}

\volumeyear{year}
\volumenumber{number}
\issuenumber{number}
\eid{identifier}
\date{\today}
\startpage{1}
\endpage{2}
\maketitle

\section{Introduction}

Three-dimensional Dirac semimetals (3DDSM) attracted a lot of attention in recent years due to their exotic electronic states. Theoretically predicted Dirac bands have also been observed by angle-resolved photoemission spectroscopy \cite{Wang2012, Young2012, Wang2013, Wehling2014}. The conduction and valence bands in 3DDSMs touch each other at the Dirac points in momentum space, and a linear energy dispersion has been shown along all momentum directions creating the three-dimensionality in these systems. Several materials have been reported such as Na$_3$Bi \cite{Wang2012}, Cd$_3$As$_2$ \cite{Neupane2014}, and SrMnBi$_2$ \cite{Park2011} with fascinating 3DDSM properties, as a 3D analogue to the 2D graphene.

The 3DDSM state is protected either by time reversal symmetry and/or inversion symmetry, and breaking one of these might drive the system into other quantum states such as topological insulators, Weyl semimetal, axion insulators, etc \cite{Krempa2014, Yang2014}. This tuning can be realized either with chemical doping and/or external effects like magnetic field, external pressure, strain, etc. These external effects are advantageous, as one can eliminate additional complications resulting from the possible impurities introduced by the chemical doping.

Among all the proposed and confirmed systems, Cd$_3$As$_2$ is one of the most studied 3DDSM. It belongs to the II$_3$-V$_2$-type narrow band semiconductors showing an inverted band structure in contrast to the other members, which made it point of interest not only today but previously, as well. Moreover, it is insensitive to the air conditions, hence easy to handle during the measurements. Due to its interesting band structure, optical \cite{Neubauer2016} and magneto-optical \cite{Akrap2016} studies at ambient pressure have been carried out, and furthermore, electrical transport and structural studies under pressure \cite{Zhang2015, He2016, Zhang2017} have been performed on this material.

 Previous pressure studies found the signatures of a pressure-induced structural phase transition between 2.5 and 4~GPa from a tetragonal $I4_1/acd$ phase to a monoclinic $P2_1/c$ phase, where the fourfold rotational symmetry is reduced to a twofold one. Even though the structural phase transition is confirmed by various studies, the pressure at which the phase transition occurs and/or an additional intermediate phase exists is still under debate. Existing studies\cite{Zhang2015, He2016, Zhang2017} show the tetragonal phase at $\sim$2.5~GPa, while the monoclinic phase is reported somewhere around 4.0~GPa. However, the lack of experimental data between 2.5 and 4.0~GPa prevents to determine the phase transition pressure precisely. This inaccuracy also affects the discussion of the relation between structural phase transition and the observed electrical transport properties, where inconsistent observations have been reported. For instance, for one of the studies \cite{Zhang2015}, the system goes into metallic-semiconducting-insulating states with increasing external pressure, where the metallic to semiconducting transition coincides with the structural phase transition at 2.5~GPa. For another study \cite{He2016}, Cd$_3$As$_2$ shows a metallic-semiconducting-metallic behavior with increasing pressure, with the metallic to semiconducting transition occuring at much lower pressures (already at 1.1~GPa) compared to the observed structural phase transition (between 2.6 and 4.67~GPa). Furthermore, for pressures above 8.5~GPa a low-temperature superconducting phase has been observed, which supports the earlier proposal of Cd$_3$As$_2$ being a candidate for topological superconductivity \cite{Wang2013}.

The inconsistent findings of earlier pressure-dependent resistivity and structural investigations motivated us to carry out a study on the charge dynamics of Cd$_3$As$_2$ under external pressure from an optical point of view. In particular, we focused on the pressure range between 2 and 4~GPa, where literature data are scarce, to elucidate possible new phases emerging under pressure.

\section{Experimental}

Large single crystals of Cd$_3$As$_2$ were synthesized with the self-selecting vapor growth method \cite{Sankar2015}. A piece with size of 250~$\times$~250~$\times$~65~$\mu$m$^3$ was used in the current measurements. Samples are polished prior to measurements and the crystallographic facet is determined as [112] surface by x-ray diffraction (XRD). The sample is placed into a type IIa diamond anvil cell (DAC) \cite{Keller1977} and finely ground CsI powder was used as a quasi-hydrostatic pressure transmitting medium. The ruby luminescence method was used for the pressure determination inside the DAC \cite{Mao1986}.

The pressure-dependent room temperature reflectivity measurements were performed with a Hyperion infrared microscope that is coupled to a Bruker Vertex 80v Fourier transform infrared spectrometer. Measurements have been done on the same crystal piece for the entire energy range ($\sim$~100-20000~cm$^{-1}$). The reflectivity spectra are measured at the sample-diamond interface, where a CuBe gasket was used as the reference. The intensity of the measured reflectivity was normalized by the intensity reflected from the CuBe gasket at the gasket-diamond interface.

The data around 2000~cm$^{-1}$ is affected by the multiphonon absorption in the diamond, therefore, this energy range was cut out and interpolated using the Drude-Lorentz fitting of the reflectivity spectra for further analysis. The optical conductivity spectra were calculated from the measured reflectivity via Kramers-Kronig (KK) transformation, as explained elsewhere \cite{Pashkin2006}.

\section{Results and Discussion}

\begin{figure}[h]%
\centering
\includegraphics[scale=9]{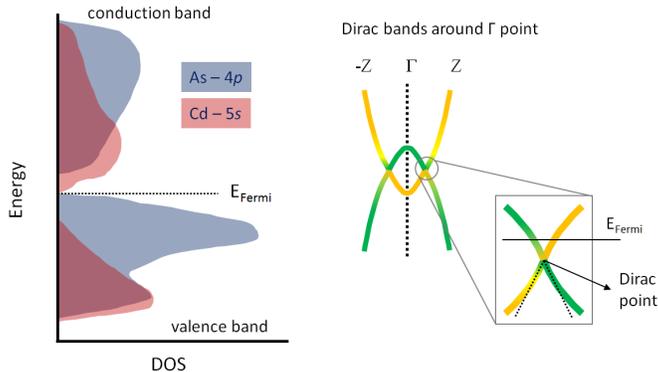}%
\caption{(color online) Schematic representation of the density of states (DOS) \cite{Wang2013} and the Dirac bands of Cd$_3$As$_2$.}%
\label{schematic}%
\end{figure}

In Fig.~\ref{schematic}, a schematic representation of the density of states (DOS)\cite{Wang2013} and the Dirac bands of Cd$_3$As$_2$ is given. The low energy electrodynamics of Cd$_3$As$_2$ is determined by As 4$d$ and Cd 5$s$ states in the vicinity of the Fermi energy. A band inversion along $\Gamma$-Z direction can be seen in the electronic structure, where one also observe Dirac bands \cite{Wang2013}. Cd$_3$As$_2$ is a naturally doped system, therefore the Fermi energy lies in the conduction band. Moreover, the deviation from the perfectly linear dispersion of the Dirac bands in Cd$_3$As$_2$ is an issue that is discussed intensively. More details regarding this issue is given in the following parts of the manuscript.

\begin{figure}[t]%
\centering
\includegraphics[scale=1]{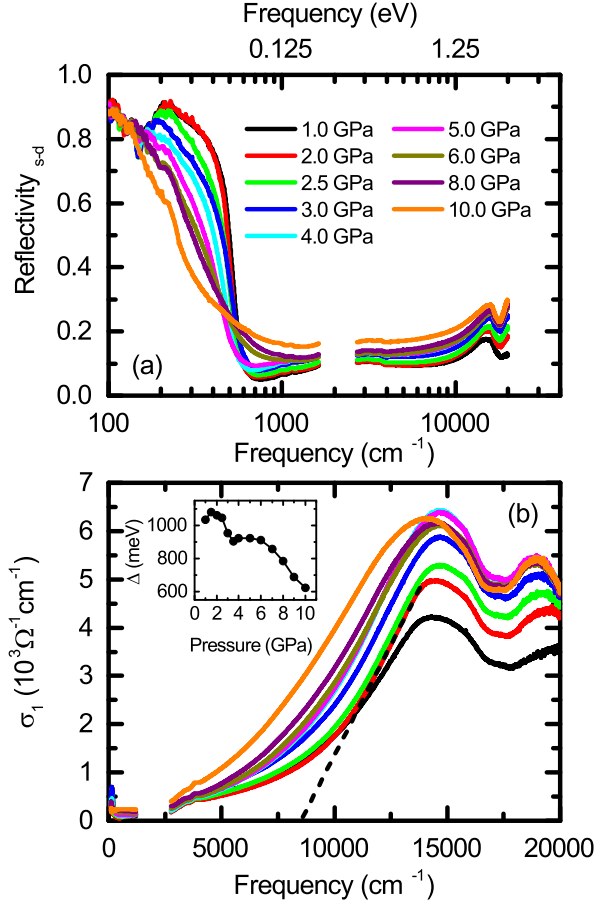}%
\caption{(color online) Pressure-dependent (a) reflectivity and (b) optical conductivity of Cd$_3$As$_2$. Inset: Pressure dependence of the high-energy band gap $\Delta$ determined by a linear extrapolation of the absorption edge to the frequency axis. An example for the linear fit (dashed line) is shown for the 2~GPa optical conductivity data.}%
\label{ref-OC}%
\end{figure}

In electrical transport measurements pronounced pressure-induced changes were observed for Cd$_3$As$_2$ under pressure \cite{Zhang2015, He2016, Zhang2017}, and therefore, one would expect to observe corresponding pressure-induced effects in optical spectra, as well.
In Fig.~\ref{ref-OC}(a) the room-temperature reflectivity spectra are depicted for selected pressures. At the lowest pressure (1.0~GPa), a sharp plasma edge is observed at $\sim$~600~cm$^{-1}$, which is an indication for a Drude-like response. This finding is also consistent with the metallic electrical resistivity shown for this material, while also expected since Cd$_3$As$_2$ is an $n$-type semiconductor, which is naturally doped. Up to 2~GPa no significant changes are observed in the reflectivity spectra. Between 2 and 5~GPa a gradual increase of the reflectivity level above the plasma edge frequency is observed, and the sharp step-like behavior due to the appearance of the plasma edge starts to be smeared out. Above 5~GPa the increase of the reflectivity level is accompanied by a sudden disappearance of the clear plasma edge.

The corresponding real part of the optical conductivity $\sigma_1(\omega)$ obtained from the reflectivity spectra via KK transformation is given in Fig.~\ref{ref-OC}(b). Up to 2~GPa an almost unchanged absorption edge at high energies is observed. Above 2~GPa a gradual shift of the absorption edge towards lower energies occurs, which is interrupted at $\sim$4~GPa, and above this pressure the continuation of the shift is visible, becoming more rapid above 8~GPa.

The interband conductivity in 3DDSM systems is expected\cite{Hosur2012, Bacsi2013} to show a specific behavior. The optical conductivity should follow a universal power-law frequency dependence according to $\sigma_1(\omega)\propto \omega^{(d-2)/z}$. Here $d$ is the dimensionality of the system and $z$ can be obtained from the band dispersion relation, $E(k)\propto|k|^z$. By using these relations, one would expect to observe an $\omega$-linear optical conductivity for a 3DDSM. Such behavior of the optical conductivity has indeed been shown for several materials \cite{Timusk2013, Chen2015}.

\begin{figure}[h]%
\centering
\includegraphics[scale=1]{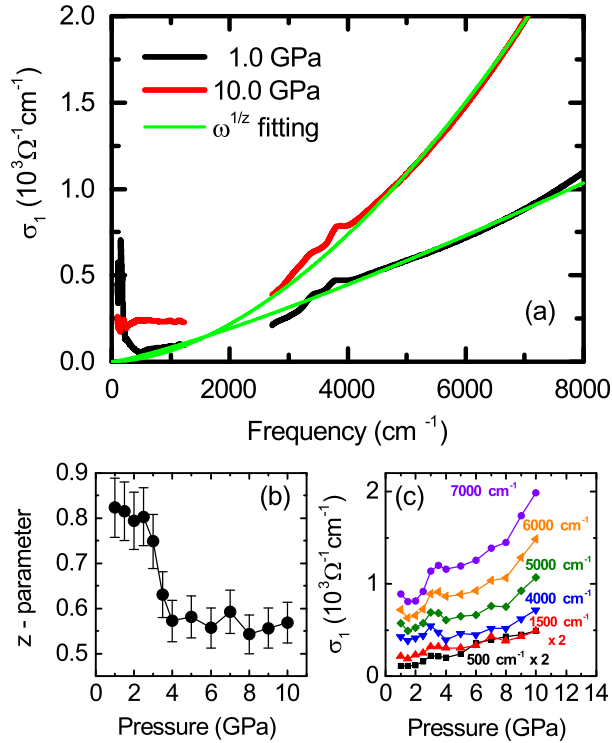}%
\caption{(color online) (a) Optical conductivity and $\omega^{1/z}$ fittings at 1.0 and 10 GPa. The small absorption features around 4000~cm$^{-1}$ are due to multi-phonon excitations in the diamond anvil. Fittings were performed in the frequency range 4000-6000~cm$^{-1}$. (b) Value of the $z$-parameter, which is a measure of the band dispersion, as a function of pressure, extracted by the fitting. (c) Pressure dependence of $\sigma_1(\omega)$ at various frequencies. The data at 500 and 1500~cm$^{-1}$ were multiplied by factor of 2 for clarity.}%
\label{z_parameter}%
\end{figure}

For Cd$_3$As$_2$, on the other hand, a sublinear dispersion will be expected, as discussed theoretically and shown by experimental probes \cite{Borisenko2014, Neupane2014, Liu2014, Jeon2014}. In the very low-energy region close to the Dirac points, it is expected that this sublinear dispersion eventually evolves to a linear one. However, it is a difficult task to observe or extract this $\omega$-linear behavior, since the existence of free charge carrier in the system shifts the Fermi energy above the Dirac point and the plasma edge behavior masks the $\omega$-linear region. More elaborate models have been proposed for doped 3DDSMs, which take into account the shift of the Fermi energy, etc., although do not account for the sublinear-dispersion \cite{Ashby2014, Tabert2016, Tabert2016b}. However, for the Cd$_3$As$_2$ samples, optical studies showed a Dirac cone Pauli-blocked edge placed around 1700~cm$^{-1}$, which implies a Fermi level around 100~meV, also consistent with the literature regarding the reported carrier concentrations \cite{Jenkins2016}. In our measurement, the energy range between 1600 and 2500~cm$^{-1}$ is affected by the multiphonon absorption of the diamond, and therefore, it is impossible for us to employ and discuss the mentioned theories, hence we analyzed our data with a simplified $\sigma_1(\omega)\propto \omega^{1/z}$ approach \cite{Timusk2013}, which we believe will show the general behavior of the pressure evolution.

In Fig.~\ref{z_parameter}(a) the fitting of the optical conductivity for 1 and 10 GPa is depicted as examples. Fittings have been performed in the frequency range 4000-6000~cm$^{-1}$ for all pressures. The small peak-like features around 4000~cm$^{-1}$ are also due to multi-phonon absorption in the diamond anvil. In Fig.~\ref{z_parameter}(b) the value of the $z$-parameter, as extracted by the fitting, is given as a function of pressure. The pressure range can be divided into two distinct regimes. In the low-pressure regime, the $z$-parameter has a value around 0.8 and does not show a significant change with increasing pressure within the error bars. Moreover, the values lower than 1 confirm the sublinear behavior of the band dispersion. With increasing pressure above 4~GPa, an abrupt drop of the $z$-parameter is observed indicating the flattening of the bands. The question arises, whether the low-pressure Dirac state survives above 4~GPa or not. The phase in the pressure range above the structural phase transition was specified as semiconducting based on the electrical transport measurements, and a band gap has been determined. However, such a gap opening at Dirac bands does not contradict with the possibility of the survival of the Dirac bands \cite{Neubauer2016, Tabert2016, Morimoto2016}.

Previous electrical transport measurements reported contradictory interpretations of the high-pressure phase in Cd$_3$As$_2$. In one of these studies \cite{Zhang2015} it has been proposed that the system undergoes a structural phase transition at around 2.5~GPa, where above this pressure the Dirac state is broken and a gradual increase of the resistivity indicates the semiconducting (and insulating with further increase of pressure) nature of the system. Another resistivity and XRD study \cite{He2016} confirmed the pressure-induced transition to a monoclinic structure and to a semiconductor-like electrical resistivity, and additionally found a metallic state above 8.5~GPa with superconductivity at low temperatures. The latter finding was discussed from the topological insulator state point of view, where one would also expect topological superconductivity. Our results do not completely exclude the conclusions drawn by either of the mentioned publications, however, a different interpretation for the high-pressure regime might be necessary.

\begin{figure}[t]%
\centering
\includegraphics[scale=1]{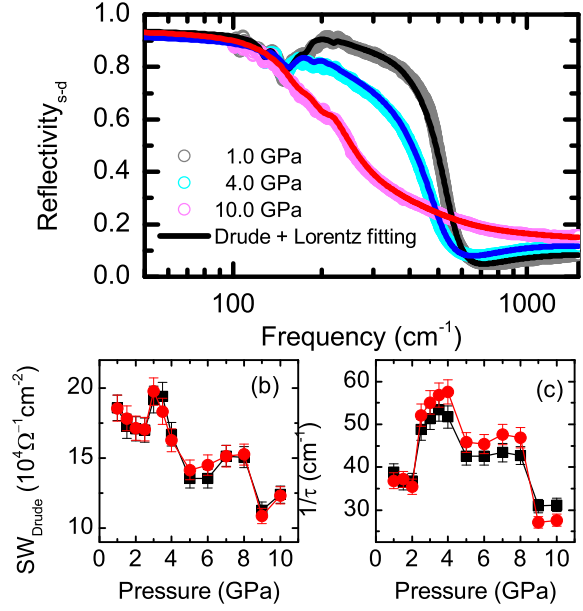}%
\caption{(color online) (a) Reflectivity and Drude+Lorentz fitting of the measured reflectivity at 1.0, 4.0, and 10~GPa. (b) Spectral weight and (c) scattering rate of the Drude component obtained from the fittings. Results from the fittings up to 1500 and 20000~cm$^{-1}$ are indicated by solid squares and solid circles, respectively.}%
\label{D_comp}%
\end{figure}

Signatures of a pressure-induced structural phase transition can also be found in the optical spectra, namely an anomaly is observed at 4~GPa in several optical parameters: (i) The abrupt drop of the $z$-parameter [Fig.~\ref{z_parameter}(b)] indicates a sudden change of the band structure, which can be due to the breakdown of the 4-fold rotational symmetry. (ii) The gradual redshift of the high-energy bandgap [see inset of Fig.\ref{ref-OC}(b)] is interrupted at 4~GPa. The latter indicates that the change happening at this pressure is sensed by the whole electronic band structure and not only by the Dirac bands, which supports the structural nature of the observed change at this pressure.
Furthermore, the shift of the high-energy bands towards lower energies (especially above 8~GPa) with an increase of the conductivity level at low energies [see Fig.~\ref{z_parameter} (c)] indicates that the system does not show an insulating behavior in contrast to Ref.\cite{Zhang2015}, but rather evolves towards a metallic state. Hence, the results of our study are more consistent with the findings in Ref.\cite{He2016}. Please note that the band gap is not completely closed, and thus a completely metallic state has not been observed within the measured pressure range.

A more detailed investigation of the low-pressure regime reveals that there might be a pressure-induced intermediate phase before the occurrence of the structural phase transition at 4~GPa. In Fig.~\ref{z_parameter}(c), the values of $\sigma_1(\omega)$ are plotted at different frequencies as a function of pressure. At each frequency an anomaly in the optical conductivity is visible between 2 and 4~GPa. To rule out the effects of the phonon modes, spectra are given in an energy range above the phonon contributions. It is a rather ambiguous process to determine the exact pressure at which the structural phase transition occurs. Previous XRD studies provide data at 2.42~GPa \cite{Zhang2015} and 2.60~GPa \cite{He2016} showing that the system is in a tetragonal phase, while the next measured pressure step was only at 3.78~GPa and 4.67~GPa, respectively, where the high-pressure monoclinic phase was reported. The large pressure step between these two phases prevents a clear determination of the phase transition pressure. This might, however, be important to clarify the existence of an intermediate phase. We point out that in Ref.~\cite{Zhang2015} a kink behavior of the Hall resistance was observed at 2.09~GPa, which might indicate the existence of an intermediate state. Moreover, a recent XRD study \cite{Zhang2017} also present data with seemingly a mixed phase at 2.92~GPa, but unfortunately a discussion of this phase has not been given.

\begin{figure}[t]%
\centering
\includegraphics[scale=1]{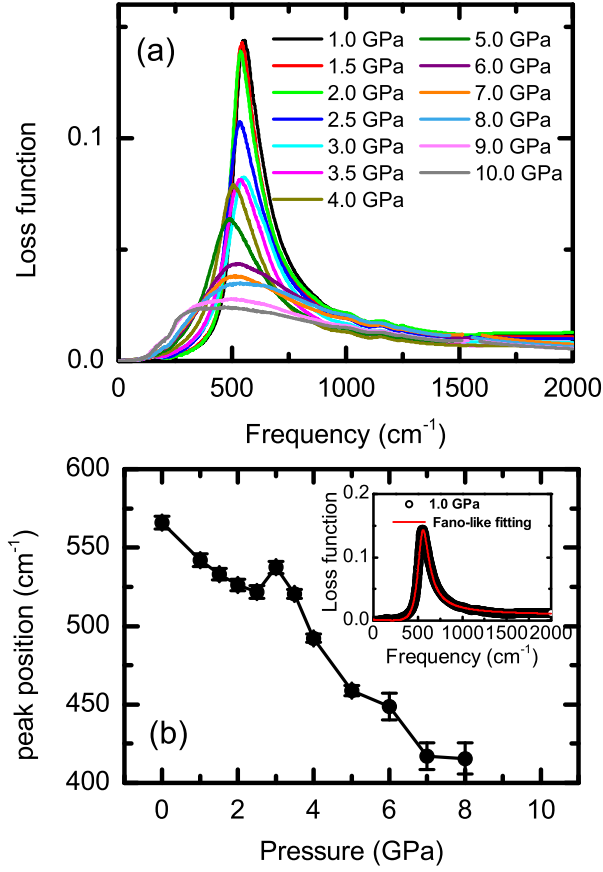}%
\caption{(color online) (a) Pressure dependence of the low-energy loss function, -Im(1/$\epsilon$($\omega$). (b) Peak position of the loss function, as obtained by the fitting. Inset: Example for the Fano-like fitting of the loss function at 1.0~GPa. }%
\label{LF}%
\end{figure}

Hints for an intermediate phase between 2 and 4~GPa can be found in other optical variables as well. In Fig.~\ref{D_comp} we present the fitting of the reflectivity at several pressures as examples [Fig.~\ref{D_comp}(a)], and the so-obtained spectral weight (SW) and scattering rate (1/$\tau$) for the observed Drude component are plotted as well [Figs.~\ref{D_comp}(b) and (c), resp.]. To obtain a better accuracy, the fittings have been performed simultaneously on reflectivity and optical conductivity data up to 1500~cm$^{-1}$, which is well above the observed plasma edge. Moreover, the fitting was repeated up to 20000~cm$^{-1}$ by taking into account the much higher-lying bands. A sudden increase of the SW is observed at 2~GPa, which could be associated with a SW transfer from high-energy interband transitions to the low-energy intraband Drude component. Such a SW transfer could be interpreted in terms of a splitting of the degenerate Dirac nodes to a Weyl semimetal phase in the framework of the Dirac fermions \cite{Tabert2016}. Concurrent with the structural phase transition the rotational symmetry in Cd$_3$As$_2$ is reduced. However, if an intermediate state exists, then it is also possible that the inversion symmetry will be disturbed by external pressure prior to the rotational symmetry lowering, which will result in a Dirac semimetal to Weyl semimetal transition. Furthermore, the observed anomalous behavior between 2 and 4~GPa might only be a precursor of the structural phase transition.

Another interesting point of the presented data is the evolution of the plasma edge, which is illustrated by the loss function (Fig.~\ref{LF}). The loss function is obtained from the dielectric function $\epsilon$($\omega$) as -Im(1/$\epsilon$($\omega$)). At the lowest pressure, the loss function shows a clear maximum at the screened plasma frequency, which confirms the plasma edge observed in reflectivity spectrum and the intraband Drude response, which is expected in this naturally doped system (Fermi energy lies in the conduction band). The width of this mode is increasing with increasing pressure and overall the mode shifts to lower energies. A Fano-like fitting \cite{Fano1961} has been performed to obtain the peak position [inset of Fig.~\ref{LF}(b)], which is given as a function of pressure in Fig.~\ref{LF}(b). Here, one notices a small jump between 2 and 4~GPa, confirming the existence of an intermediate phase as revealed by anomalies in various optical parameters as discussed above. With further pressure increase, it seems that the high-energy contributions superpose the clear plasma behavior, however, the existence of the free carriers is clearly revealed in the reflectivity spectra by the Hagen-Rubens upturn towards lower frequencies. At these high pressures the clear peak-like structure in the loss function disappears, and hence the peak position above 8~GPa is not discussed. This smearing out of the plasmon mode with increasing pressure is another evidence that the high-energy interband transitions are pushed towards the lower-energy region and start to overlap with the low energy excitations, indicating the increasing metallicity of the system.

The evolution of several optical parameters at the critical pressure of 8~GPa is also intriguing. Although no further structural transitions have been reported for this pressure range, we observe several anomalies in the optical spectra. Namely, (i) the drop of the Drude SW and the scattering rate [Fig.~\ref{D_comp}(b) and (c)] followed by the increasing trend of the Drude SW, (ii) a more rapid increase of the optical conductivity at various energy ranges [Fig.\ref{z_parameter}(c)], and (iii) the broadening and disappearance of the plasmon peak in the loss function [(Fig.~\ref{LF})(a)]. While all these changes indicate that the system goes into a more metallic state above 8~GPa, we would like to point out that this pressure range coincides with the range, where low-temperature superconductivity started to be observed in electrical transport measurements \cite{He2016}.

\section{Conclusions}

In conclusion, our pressure-dependent optical study of the 3DDSM Cd$_3$As$_2$ confirm the occurrence of a structural phase transition at 4~GPa in terms of anomalies in the optical spectra: (i) The $z$-parameter of the band dispersion relation ($E(k)\propto|k|^z$) shows an abrupt drop. (ii) The redshift of the high-energy band is interrupted. For pressures above 4~GPa the continuous redshift of the high-energy bands indicates that system goes towards a metallic state, however, a complete transition could not be observed within the measured pressure range. Moreover, the existence of Hagen-Rubens-like reflectivity with the disappearance of the plasmon peak indicates the shift of the high energy bands towards lower energies, which gives further evidence for the increasing metallicity of the system. For pressures between 2and 4 GPa we observe anomalies in the optical response, for example in the Drude parameters, which suggest the existence of an intermediate state, with a possible pressure-induced splitting of the Dirac nodes and a phase transition to a Weyl semimetal state. For the clarification of the nature of this intermediate phase detailed structural studies in this pressure range are necessary. We find anomalies in the optical response at 8~GPa, where the emergence of a low-temperature superconducting phase was observed.

\section{Acknowledgments}

We thank A. A. Tsirlin and J. Carbotte for fruitful discussions. This project is financially supported by the Federal Ministry of Education and Research (BMBF), Germany, through Grant No. 05K13WA1 (Verbundprojekt 05K2013, Teilprojekt 1, PT-DESY).

$^*$ ece.uykur@physik.uni-augsburg.de

$^\dagger$ christine.kuntscher@physik.uni-augsburg.de

\bibliographystyle{apsrev4-1}
\bibliography{References}

\end{document}